\documentclass[12pt]{article}
\newcommand{\bm}[1]{\mbox{\boldmath $#1$}}

\def\fn{\mbox{$f_{0}$(980)}}
\def\ft{\mbox{$f_{0}$(1370)}}
\def\ff{\mbox{$f_{0}$(1500)}}

\def\kf{\mbox{$K_0^*$(1430)}}
\def\an{\mbox{$a_0$(980)}}
\def\af{\mbox{$a_0$(1450)}}
\def\Dfp{\mbox{$D_{s}^{+}\rightarrow f_{0}(980)\pi^{+}$ }}

\begin{document} \baselineskip .7cm
\title{Comment on ``Intrinsic and dynamically generated scalar meson states''}
\author{
George Rupp\\
{\normalsize\it Centro de F\'{\i}sica das Interac\c{c}\~{o}es Fundamentais}\\
{\normalsize\it Instituto Superior T\'{e}cnico, Edif\'{\i}cio Ci\^{e}ncia}\\
{\normalsize\it P-1096 Lisboa Codex, Portugal}\\
{\small george@ajax.ist.utl.pt}\\ [.3cm]
Eef van Beveren\\
{\normalsize\it Centro de F\'{\i}sica Te\'{o}rica}\\
{\normalsize\it Departamento de F\'{\i}sica, Universidade de Coimbra}\\
{\normalsize\it P-3000 Coimbra, Portugal}\\
{\small eef@teor.fis.uc.pt}\\ [.3cm]
\and
Michael D. Scadron\\
{\normalsize\it Physics Department}\\
{\normalsize\it University of Arizona}\\
{\normalsize\it Tucson, AZ, 85721 USA}\\
{\small scadron@physics.arizona.edu}\\ [.3cm]
{\small PACS number(s): 14.40.Cs, 12.39.Pn, 13.75.Lb}\\ [.3cm]
{\small hep-ph/0104087}
}
\date{\today}
\maketitle

\begin{abstract}
The scalar-meson assignments of Shakin and Wang in a generalized
Nambu--Jona-Lasinio model are contradicted by recent experimental information.
Also the strict distinction made by these authors between ``intrinsic'' and
``dynamically generated'' states is contested, as well as a number of other
statements.
\end{abstract}

In Ref.~\cite{SW00D}, Shakin and Wang (SW) revisit a generalized
Nambu--Jona-Lasinio (NJL) model, recently applied to light \cite{CHWS99} and
scalar \cite{CGHWS00} mesons, so as to present what the authors claim to be
additional evidence for their model assignments of scalar-meson resonances.
Essential for the interpretation of scalar mesons in SW's model is the
distinction between what they call ``intrinsic''  or ``preexisting'' (IP), and
``dynamically generated'' (DG) scalar states, only the former ones
corresponding to $q\bar{q}$ quark-model states that should form nonets. In
contrast, the DG states, not necessarily forming nonets, are supposed to be the
result of $t$- and $u$-channel meson exchange in $S$-wave meson-meson
scattering, giving rise to the $\sigma$(500--600) (or $f_0$(400--1200))
and, together with threshold effects in the $q\bar{q}$ $T$~matrix, also the
--- not yet established --- $\kappa$(900) (or $K_0^*$(700--1100)). In this
Comment, we want to point out that not only is there quite compelling
experimental evidence against some of the assignments of SW, but also their
strict distinction between IP and DG is a model-dependent simplification which
may be quite misleading.

Starting with the assignments, SW right away present a dubious argument against
placing the \af\ and the \kf\ in the same nonet, arguing that one would expect
the \kf\ to be more massive than the \af. While this could be true in a very
naive quark-model picture, it is very dangerous to apply such a line of
reasoning to broad resonaces like the scalar mesons under consideration, which
are subject to strong unitarization effects, as advocated by e.g.\ Maltman
\cite{M99}, who is extensively quoted by SW.
Moreover, by the same token one could argue that the \ft, which is interpreted
by SW as an $s\bar{s}$ state lying in the same nonet as the \kf, should be the
more massive one. Clearly, naive arguments are inadequate to understand the 
scalars, as the large mass shifts for the $f_0$s and $a_0$s in SW's work, due
to a short-range NJL interaction, already indicate. Let us just add to this
point that it seems much more natural and appealing to place all the scalars
below 1 GeV in one nonet, which was accomplished by us in previous work
\cite{S8284,BRMDRR86}, as well as by several other authors 
\cite{TR96a,IIITT97,J77S80,OOP99}, besides Schechter and co-workers (see e.g.\
Ref.~\cite{BFSS99} and references in \cite{SW00D} quoted by SW). In this
picture, the scalar mesons between say 1.3 and 1.5 GeV belong to another nonet,
and so forth.  If this can be achieved by unitarization only without having to
resort to rather {\em ad hoc} \/interactions besides the confinement mechanism,
which is indeed the case in the unitarized meson model (denoted by NUMM) of two
of us \cite{BRMDRR86,BR99}, all the better.

Let us now analyse in more detail the interpretation SW attribute to some
well-established scalar mesons, i.e., the \fn, \ft, and \ff. \\[2mm]
\bm{f_{0}(980):} \\[1mm]
This isoscalar scalar meson is described by SW as the lowest $n\bar{n}$
state. However, as early as in 1989, the DM2 collaboration \cite{DM89} not only
confirmed the $\sigma$ meson as an $S$-wave two-pion resonance in the decay
process $J\Psi\rightarrow\omega\pi\pi$, with much higher statistics than the 
equivalent Mark-1 experiment over a decade earlier \cite{Ma77}, but also
produced a
clear indication that the \fn\ is {\em not} \/mainly $n\bar{n}$. The point is
that in the same $\pi\pi$ mass distribution where a huge $\sigma$ bump shows
up only a tiny \fn\ peak is observed, hinting at a dominant $s\bar{s}$ 
structure for this resonance.
But also
the very recently measured weak decay rate \cite{PDG00}
\begin{equation}
\Gamma(\Dfp) = (2.39\pm1.06) \times 10^{-14}\,\; \mbox{GeV}
\end{equation}
is clear evidence for the \fn\ being mostly $s\bar{s}$, since we have shown
\cite{BRS00} that this rate can be pefectly reproduced through a standard
$W^+$-emission process (see Figure~1), provided one assumes a dominantly
\begin{figure}[ht]
\Large
\begin{center}
\begin{picture}(300,150)(-150,-100)
\put(-100,-50){\line(1,0){200}}
\put(-100,-90){\line(1,0){200}}
\put(4,-5){\line(1,0){96}}
\put(4,-5){\line(4,1){96}}
\put(-102,-50){\makebox(0,0)[rc]{$c$}}
\put(102,-50){\makebox(0,0)[lc]{$s$}}
\put(-102,-90){\makebox(0,0)[rc]{$\bar{s}$}}
\put(102,-90){\makebox(0,0)[lc]{$\bar{s}$}}
\put(-20,-22){\makebox(0,0)[rc]{$W^{+}$}}
\put(102,-5){\makebox(0,0)[lc]{$\bar{d}$}}
\put(102,19){\makebox(0,0)[lc]{$u$}}
\put(-125,-70){\makebox(0,0)[rc]{$D_{s}^{+}$}}
\put(125,-70){\makebox(0,0)[lc]{$f_{0}(s\bar{s})$}}
\put(125,7){\makebox(0,0)[lc]{$\pi^{+}$}}
\end{picture}
\end{center}
\normalsize
\caption[]{Contribution of $W^{+}$ emission to the weak decay $D_s^+\rightarrow
f_0(s\bar{s})\pi^+$.}
\label{Wemission}
\end{figure}
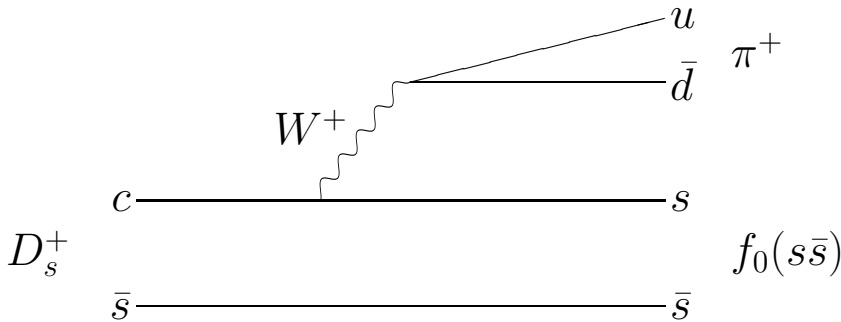
$s\bar{s}$ configuration for the \fn, possibly with a small $n\bar{n}$
admixture.
\bm{f_{0}(1370):} \\[1mm]
Although the different hadronic branching fractions of this resonance are not
very well known experimentally, the dominant decay modes involve two and
four pions \cite{PDG00}, indicating that the \ft\ is mostly $n\bar{n}$. In
particular, the available data give a branching ratio $\Gamma(K\bar{K})/
\Gamma_{\mbox{\scriptsize total}} = 0.35 \pm 0.13$ \cite{B96}, which is
in accordance with a mainly nonstrange \ft, while SW classify it as an
$s\bar{s}$ state. Also very recent data support our $n\bar{n}$ interpretation
of the \ft, namely the failure to observe the process
$D_s^+\rightarrow\ft\pi^+\rightarrow K^+K^-\pi^+$ \cite{E791} (see also
\cite{BRS00}), and the dominance of $J/\Psi\rightarrow\phi\ft\rightarrow
\phi\pi\pi$ over $J/\Psi\rightarrow\phi\ft\rightarrow \phi K\bar{K}$
\cite{B01}. \\[2mm]
\bm{f_{0}(1500):} \\[1mm]
For this resonance, we can again apply the $W^+$-emission graph of Fig.~1,
since the weak decay $D_s^+\rightarrow \ff\pi^+$ has been observed very
recently, with the rate $(3.7 \pm 2.1) \times 10^{-15}$ GeV \cite{PDG00}. If
we assume a pure $s\bar{s}$ configuration for the \ff, we obtain a
theoretical decay rate of $3.3 \times 10^{-15}$ GeV. Of course, the large
experimental error bar perfectly allows for some $n\bar{n}$ admixture in the
\ff, but a pure $n\bar{n}$ assignment as advocated by SW seems highly unlikely.
As for the hadronic decays of the \ff, the different branching fractions are
even less well known than in the \ft\ case. Nevertheless, the dominant decay
modes of the \ff\ involve $\eta$s and $\eta\prime$s, having non-zero
strange-quark contents, and not pions like for the \ft, which also hints at a
dominant $s\bar{s}$ structure for the \ff. At this point we should mention that
the relatively small width of the \ff\ and its tiny branching fraction into
$K\bar{K}$ \cite{B96} are often invoked as being evidence for a glueball
interpretation of this resonance. However, these peculiar properties can be
understood instead by assuming the \ff\ to be close to a flavor-octet
configuration, but still dominantly $s\bar{s}$. This would give rise to
destructive interference between the $s\bar{s}$ and $n\bar{n}$ components,
leading to a strong suppression of the $K\bar{K}$ mode \cite{KMMP95,BR99},
which would be among the dominant decay modes if the \ff\ was purely
$s\bar{s}$. As we have observed above, the large experimental error for the
weak decay $D_s^+\rightarrow \ff\pi^+$ can easily accomodate a significant
$n\bar{n}$ admixture in the \ff, so that the octet hypothesis is  plausible.

Summarizing, the experimental data do not favor SW's $s\bar{s}$ and $n\bar{n}$
assignments for the \ft\ and \ff, respectively, nor do the recent lattice
calculations of Lee and Weingarten \cite{LW00} for that matter, as mentioned by
SW.

Now we turn to the question of IP versus DG scalar states. This is in fact not
a new issue, and has already been explicitly addressed by e.g.\ Isgur \& Speth
(IS), in a Comment \cite{IS96} on the work of T{\"{o}rnqvist \& Roos (TR)
\cite{TR96a}. Though disagreeing with SW on the nature of the \an\ and \fn,
also IS argue that  light scalars owe their existence
to ``degrees of freedom already present in the meson-meson continuum'', i.e.,
$t$-channel forces, to be contrasted with ``intrinsic poles arising from the
insertion of a new $q\bar{q}$ degree of freedom''. At the same time, IS
criticize TR for the omission of $t$-channel meson exchanges, which according
to them calls into question TR's analysis. However, in another Comment on the
same paper, Harada {\em et al.} \cite{HSS97} quantitatively demonstrate, in
the framework of their own model, that the neglect of $\rho$-meson exchange
in the $S$-wave $\pi\pi$ amplitude, though destroying crossing symmetry, does
not destroy the existence of the $\sigma$ meson, and does not even worsen the
quality of the fit, only leading
to a moderate (complex) shift of the $\sigma$ pole. This finding lends
support to TR's claim, seconded by us, in their Reply \cite{TR96b} to IS that
``a detailed inclusion of all nearby $s$-channel singularities is more 
important than the inclusion of a few strong $t$-channel exchanges'' (see also
Ref.~\cite{S99} for further discussion on the $\sigma$, crossing, and chiral
symmetry).

We wish to add to this discussion by arguing that the strict separation of 
IP and DG poles, as advocated by IS and SW, is a model artifact, which is
probably a much more serious approximation than the neglect of $t$-channel
exchanges in the NUMM \cite{BRMDRR86,BR99} and the model of TR \cite{TR96a}.
The crucial point is that, once one accepts strong three-meson couplings, as IS
and SW seem to do, these will inexorably show up also in the scalar
$\rightarrow$ pseudoscalar-pseudoscalar (and scalar $\rightarrow$
vector-vector) sector. Hence any ``intrinsic'' scalar state will couple
strongly to the ``meson-meson continuum'', leading to large unitarization
effects. This will inevitably give rise to strong mixing of IP and DG states,
making a strict identification of either type somewhat meaningless. (However,
pure DG chiral schemes at the quark level which involve scalar mesons do appear
to have merit \cite{DS9598}.) In the
NUMM, which is a coupled-channel model where the $q\bar{q}$ and meson-meson
sectors are treated on an equal footing, unitarization leads to a phenomenon
unique to scalar mesons, namely resonance doubling, also observed by TR. So
even without including $t$-channel exchanges extra poles are generated, which
can be interpreted as the light scalars and, moreover, allow a reasonably
good decription, without any free parameters, of $S$-wave meson-meson phase
shifts up to about 1.2 GeV in the case of the NUMM \cite{BR99,BRMDRR86}. But
this does not mean that the poles below 1 GeV are of a DG nature, while the
ones above 1 GeV are of the IP type. It namely happens that either set of poles
can be traced back to the ``intrinsic'' $q\bar{q}$ bound states (see
Ref.~\cite{BR99} for more details and references).

Finally, we discuss the meson decay constants SW invoke
as apparent support for their scalar-meson assignments. While we do not have
any fundamental objection against such a procedure, one should realize that
these constants are not observables and, therefore, model dependent.
However, we note that SW compare their decay constants with those of Maltman
\cite{M99}, who claims that the values he finds for the \an, the \af, and the
\kf\ ``suggest a UQM-like (unitarized quark model) scenario for the isovector
scalar states'',  a scenario which is clearly not considered by SW, with one
exception. Since SW find disagreement with Maltman's value in the \an\ case,
they admit ``\ldots which suggests that the \an\ may have a significant
$K\bar{K}$ component''. But then they should allow the inclusion of sizable
two-meson components also in the case of the other scalar mesons, as naturally
happens in UQMs like the NUMM and the model of Ref.~\cite{TR96a}. This
could considerably change the results for the decay constants.

In conclusion, we believe to have demonstrated in this Comment that the
interpretation of scalar-meson states by SW is clearly called into question by
experiment. Furthermore, their strict distinction between IP and DG scalar
states lacks a consistent theoretical foundation.
\vspace{0.3cm}

{\bf Acknowledgement}:
This work is partly supported by the
{\it Funda\c{c}\~{a}o para a Ci\^{e}ncia e a Tecnologia}
of the {\it Minist\'{e}rio da
Ci\^{e}ncia e da Tecnologia} \/of Portugal,
under contract numbers
PESO/\-P/\-PRO/\-15127/\-99,
POCTI/\-35304/\-FIS/\-2000,
and
CERN/\-P/\-FIS/\-40119/\-2000.

\end{document}